\documentclass[letter]{emulateapj}

\usepackage{natbib}
\usepackage{amssymb}
\usepackage{amsmath}
\usepackage{color}
\usepackage{enumitem}
\usepackage{graphicx}

\definecolor{myred}{rgb}{0.8,0.0,0.0}

\newcommand{\rxj}{RX~J1347.5$-$1145}
\newcommand{\vpec}{$v_{\textrm{pec}}$}
\newcommand{\rfive}{$R_{2500}$}
\newcommand{\bplusm}{Bolocam$+$MUSIC}
\newcommand{\mjy}{MJy~sr$^{-1}$}
\newcommand{\taue}{$\tau_{\mathrm{e}}$}
\newcommand{\kms}{km~s$^{-1}$}
\newcommand{\spire}{{\it Herschel}--SPIRE}
\newcommand{\Te}{$T_{\mathrm{e}}$}
\newcommand{\chandra}{{\it Chandra}}
\newcommand{\mgas}{$M_{\textrm{gas}}$}

\begin{document}

\title{Peculiar Velocity Constraints from Five-Band
  SZ Effect Measurements Towards \rxj\ with MUSIC and Bolocam from the CSO}

\author{
    Jack~Sayers\altaffilmark{1,12},
    Michael~Zemcov\altaffilmark{2},
    Jason~Glenn\altaffilmark{3},
    Sunil~R.~Golwala\altaffilmark{1},
    Philip~R.~Maloney\altaffilmark{3},
    Seth~R.~Siegel\altaffilmark{1}, 
    Jordan~Wheeler\altaffilmark{3},
    Clint~Bockstiegel\altaffilmark{4},
    Spencer~Brugger\altaffilmark{3},
    Nicole~G.~Czakon\altaffilmark{5},
    Peter~K.~Day\altaffilmark{6},
    Thomas~P.~Downes\altaffilmark{7},
    Ran~P.~Duan\altaffilmark{8},
    Jiansong~Gao\altaffilmark{9},
    Matthew~I.~Hollister\altaffilmark{1},
    Albert~Lam\altaffilmark{1},
    Henry~G.~LeDuc\altaffilmark{6},
    Benjamin~A.~Mazin\altaffilmark{4},
    Sean~G.~McHugh\altaffilmark{4},
    David~A.~Miller\altaffilmark{1},
    Tony~K.~Mroczkowski\altaffilmark{10},
    Omid~Noroozian\altaffilmark{11},
    Hien~T.~Nguyen\altaffilmark{6},
    Simon~J.~E.~Radford\altaffilmark{1},
    James~A.~Schlaerth\altaffilmark{1,3},
    Anastasios~Vayonakis\altaffilmark{1},
    Philip~R.~Wilson\altaffilmark{6},
    \& Jonas~Zmuidzinas\altaffilmark{1}
 }
 \altaffiltext{1}
   {Division of Physics, Math, and Astronomy, California Institute of Technology, 
     Pasadena, CA 91125}
 \altaffiltext{2}
   {Rochester Institute of Technology, Rochester, NY 14623}
 \altaffiltext{3}
   {Department of Astrophysical and Planetary Science,
     University of Colorado, Boulder, CO 80309}
 \altaffiltext{4}
   {Department of Physics, University of California, Santa Barbara, CA 93106}
 \altaffiltext{5}
   {Institute of Astronomy and Astrophysics, Academia Sinica, 
      Taipei, Taiwan}
 \altaffiltext{6}
   {Jet Propulsion Laboratory, Pasadena, CA 91109}
 \altaffiltext{7}
   {Department of Physics, University of Wisconsin, Milwaukee, WI 53201}
 \altaffiltext{8}
   {National Astronomical Observatories, Chinese Academy of Sciences,
     Chaoyang District, Beijing, China}
 \altaffiltext{9}
   {National Institute of Standards and Technology, Boulder, CO 80305}
 \altaffiltext{10}
   {National Research Council Fellow, United States Naval Research Laboratory, Washington, DC 20375}
 \altaffiltext{11}
   {Goddard Space Flight Center, Greenbelt, MD 20771}
 \altaffiltext{12}
    {jack@caltech.edu}

\begin{abstract}

  We present Sunyaev-Zel'dovich (SZ) effect measurements from wide-field images
  towards the galaxy cluster \rxj\ obtained from the
  Caltech Submillimeter Observatory with
  the Multiwavelength Submillimeter Inductance Camera (MUSIC) at 147, 213, 281, and 337~GHz
  and with Bolocam at 140~GHz.
  As part of our analysis, we have used higher frequency data from \spire\
  and previously published lower frequency radio data to subtract the signal from the brightest
  dusty star-forming galaxies behind \rxj\ and from the AGN in \rxj's BCG.
  Using these five-band SZ effect images, combined with 
  X-ray spectroscopic measurements of the temperature of the intra-cluster medium (ICM)
  from \chandra, we constrain the ICM optical
  depth to be $\tau_{\mathrm{e}} = 7.33^{+0.96}_{-0.97} \times 10^{-3}$ and the ICM
  line of sight peculiar velocity to be $v_{\mathrm{pec}} = -1040^{+870}_{-840}$~\kms.
  The errors for both quantities are limited by measurement noise rather than calibration
  uncertainties or astrophysical contamination, and significant improvements
  are possible with deeper observations.
  Our best-fit velocity is in good agreement with one previously published SZ effect
  analysis and in mild tension with the other, although some or all of that
  tension may be because that measurement samples a much smaller cluster volume.
  Furthermore, our best-fit optical depth implies a gas mass slightly larger
  than the \chandra-derived value, implying the cluster is elongated along
  the line of sight.

\end{abstract}

\section{Introduction}

  The peculiar velocities (\vpec) of galaxy clusters and other large-scale structures
  are related to the underlying cosmology, particularly the properties
  of dark energy and the long length-scale behavior of gravity.
  As such, measurements of \vpec\ can be used to place constraints
  on a variety of cosmological parameters (e.g., \citealt{Kaiser1987, 
    Scoccimarro2004, Bhattacharya2008, Kosowsky2009}).
  In addition, the assumption of large-scale homogeneity in the universe
  based on the Copernican principle can also be tested using \vpec\ measurements
  \citep{Clarkson2008, Clarkson2012, Planck2014_XIII}.
  The kinetic Sunyaev-Zel'dovich (SZ) effect, which describes the Doppler shift
  of Cosmic Microwave Background (CMB) photons that scatter off of hot electrons in the intra-cluster medium (ICM),
  provides an excellent tool for such measurements \citep{Sunyaev1972, Birkinshaw1999, Carlstrom2002}.
  In particular, the surface brightness of the kinetic SZ effect signal is independent of redshift
  and is directly proportional to the line of sight peculiar velocity of the ICM relative
  to the fixed reference frame of the CMB.

  However, the relative dimness of the kinetic SZ effect signal, along with
  contamination from several other astrophysical sources such as primary
  CMB fluctuations and the thermal SZ effect, present a significant
  observational challenge. Recently, progress has been
  made by averaging the kinetic SZ effect signal over large samples
  of objects in wide field CMB surveys.
  For example, kinetic SZ effect measurements based on
  multi-band data from the {\it WMAP} and {\it Planck} 
  satellites have been used to place upper limits on the average and 
  rms peculiar velocities of large ($\simeq 1000$) X-ray selected cluster samples
  \citep{Kashlinsky2008, Osborne2011, Planck2014_XIII}. 
  In addition, {\it Planck} and the ground-based Atacama Cosmology Telescope (ACT) 
  have made separate modest-significance detections of the
  kinetic SZ effect signal using pairwise stacks on galaxies
  detected in the Sloan Digital Sky Survey \citep{Planck2015_XXXVII, Hand2012}.
  Furthermore, data from {\it Planck}, ACT, and the ground-based
  South Pole Telescope (SPT) imply a non-zero kinetic SZ effect
  power spectrum on small angular scales \citep{Planck2015_XI, George2015, Sievers2013}.
  Complementary to these large statistical analyses, 
  several efforts have also been made to detect the kinetic SZ effect signal 
  in individual clusters using multi-band ground-based measurements
    \citep{Benson2003, Kitayama2004,
    Mauskopf2012, Zemcov2012, Mroczkowski2012, Lindner2015, Adam2015},
  with an exceptionally high velocity merging component in the cluster
  MACS J0717.5$+$3745 producing the highest significance detection to date
  \citep{Sayers2013_3}.

  This analysis focuses on SZ effect measurements towards the cluster \rxj, 
  which is a well-studied system that
  was the target of some of the earliest published thermal SZ effect detections
  \citep{Komatsu1999, Pointecouteau1999, Komatsu2001, Pointecouteau2001, Reese2002}.
  Due to its brightness, \rxj\ has also been used for the initial studies published 
  from several recent SZ effect instruments \citep{Mason2010, Zemcov2012, Adam2014}.
  In addition, two separate kinetic SZ effect measurements towards \rxj\
  have been published.
  \citet{Kitayama2004} found $v_{\mathrm{pec}} = -1420^{+1170}_{-1270}$~\kms\
  based on 21, 150, and 350~GHz measurements collected from the Nobeyama 45~m
  Radio Telescope and the
  James Clerk Maxwell Telescope, and
  \citet{Zemcov2012} used high spectral resolution 200--300~GHz data from
  Z-Spec combined with 140~GHz Bolocam data 
  to constrain $v_{\mathrm{pec}} = +450 \pm 810$~\kms.\footnote{
    \citet{Kitayama2004} report a value of $+1420$~\kms, but that value is based on
    an opposite convention for the velocity direction compared to what is used
    in this work and in \citet{Zemcov2012}.}

  In this manuscript we present SZ effect measurements towards \rxj\ collected
  at 147, 213, 281, and 337~GHz using the
  Multiwavelength Submillimeter Inductance Camera (MUSIC) and 
  at 140~GHz with Bolocam. Summaries of the instrument characteristics
  are given in Section~\ref{sec:instrument}, and 
  the data reduction procedures are described in Section~\ref{sec:reduction}.
  Section~\ref{sec:X-ray} details 
  the analysis of \chandra\ X-ray exposures to obtain a spectroscopic constraint
  on the intra-cluster medium (ICM) temperature \Te.
  Section~\ref{sec:model} discusses the model used to describe the shape
  of the SZ effect signal towards \rxj, and Section~\ref{sec:spectral_fit}
  details how the SZ brightness measurements derived from the model
  fits, in combination with the X-ray spectroscopic measurements,
  are used to constrain the ICM optical depth \taue\ and peculiar
  velocity \vpec.
  A comparison to previously published results is
  given in Section~\ref{sec:discussion}.
  Throughout this manuscript, all physical quantities have been derived
  using a flat cosmology with $\Omega_{\mathrm{M}} = 0.3$, $\Omega_{\lambda} = 0.7$,
  and $H_0 = 70$~\kms~Mpc$^{-1}$.

\section{SZ Effect Instrumentation}
  \label{sec:instrument}

  \subsection{MUSIC}
    \label{sec:music}
 
    MUSIC was built to serve as the long-wavelength imaging camera for
    the Caltech Submillimeter Observatory (CSO), and its overall design and construction
    is detailed in \citet{Glenn2008}, \citet{Maloney2010}, \citet{Golwala2012}, 
    and \citet{Sayers2014}.
    Relative to the instrument description in \citet{Sayers2014}, 
    the cryogenic lens has been redesigned in order to change the plate scale
    of the focal plane from $7''$/mm to $11.5''$/mm. As originally
    built, eight focal plane tiles would fill MUSIC's $14'$ circular
    field of view (FOV). However, the initial fabrication yielded
    only two working tiles, and this plate scale change allows these
    two tiles to fill a square $11.5'$ FOV ($14'$ diagonal).
    Subsequent to this redesign, one of the working
    tiles was damaged, and all of the data described in this
    manuscript were collected using a single tile with an
    approximate FOV of $5.8' \times 11.5'$ (AZ~$\times$~EL). 

    The MUSIC focal plane uses phased arrays of slot antennas to couple to
    incoming radiation over a broad bandwidth spanning from
    $\simeq 100$--$400$~GHz \citep{Goldin2003}. The output from each
    antenna array is then coupled to four separate microwave kinetic
    inductance detectors (MKIDs), with three-pole lumped element
    filters defining a different observing band for each MKID
    \citep{Kumar2009, Duan2015}.
    These four observing bands are centered
    at 147, 213, 281, and 337~GHz,\footnote{
      These values represent the effective band centers for a 
      relativistically corrected thermal
      SZ spectrum assuming an ICM temperature of 10~keV.}
    and they have corresponding point spread
    functions (PSFs) that are approximately Gaussian and 
    have full-width half maxima (FWHM) of 
    $48''$, $36''$, $32''$, and $29''$.\footnote{These FWHM are 
    slightly larger than the values given in \citet{Sayers2014}
    due to the optical reconfiguration.}
    Each MUSIC focal plane tile consists of 72 antenna arrays and 288 MKIDs,
    and the tile used to collect the data described in this manuscript
    has 200 MKIDs that pass all of the initial quality checks.
    An additional cut is made based on the on-sky sensitivity of each 
    detector, removing all MKIDs with a sensitivity more
    than two standard deviations poorer than the median sensitivity within
    each observing band.
    After this cut 125 MKIDs remain, with 30, 35, 32, and 28 MKIDs
    at 147, 213, 281, and 337~GHz.

  \subsection{Bolocam}
    \label{sec:bolocam}

    Bolocam served as the long-wavelength imaging camera at the CSO prior
    to the installation of MUSIC, and the instrument is described in 
    detail in \citet{Glenn1998} and \citet{Haig2004}.
    Bolocam operated with an $8'$ circular FOV filled with 144 
    spiderweb bolometers using neutron-transmutation-doped Ge thermistors, 
    of which approximately 115 were optically coupled.
    The data described in this manuscript were collected with
    Bolocam operating at 140~GHz, where the PSFs had a FWHM of $58''$.
 
    As detailed in Section~\ref{sec:spectral_fit}, the Bolocam data
    provide a modest improvement in constraining the SZ effect
    signal compared to using the MUSIC data alone.
    In addition, this is the first SZ analysis using MUSIC data,
    while the Bolocam data for \rxj\ have already been published
    in multiple analyses (e.g., \citealt{Sayers2013_1, Sayers2013_2, Czakon2015}).
    As a result, the well-vetted Bolocam data provide a useful consistency check
    for the MUSIC measurements.

\section{SZ Data Reduction}
 \label{sec:reduction}

  \subsection{General Reduction and Calibration}
    \label{sec:cal}

    \begin{deluxetable*}{cccccc}
      \tablewidth{0pt}
      \tablecaption{\bplusm\ Data}
      \tablehead{\colhead{band (GHz)} & \colhead{PSF} & \colhead{HPF (half-signal)} & 
        \colhead{HPF ($2 \times $\rfive)} & \colhead{rms/beam} & \colhead{image size}}
      \startdata 
        140 GHz & $58''$ & $(9.5')^{-1}$ & 0.89 & 1.1 mJy & $14' \times 14'$ \\
        147 GHz & $48''$ & $(2.5')^{-1}$ & 0.30 & 1.1 mJy & $12' \times 12'$ \\
        213 GHz & $36''$ & $(2.4')^{-1}$ & 0.26 & 0.8 mJy & $12' \times 12'$ \\
        281 GHz & $32''$ & $(2.9')^{-1}$ & 0.33 & 2.0 mJy & $12' \times 12'$ \\
        337 GHz & $29''$ & $(2.7')^{-1}$ & 0.29 & 7.2 mJy & $12' \times 12'$ 
      \enddata
      \tablecomments{The \bplusm\ data used in this analysis. The columns
        give the thermal SZ effect-weighted band center, the PSF FWHM, 
        the angular scale at which half of the signal from a cluster-sized
        object is attenuated due to the effective high-pass filtering
        of the data processing, the fraction of signal remaining for angular scales
        corresponding to $2 \times $\rfive\ = 4.1$'$ (see Section~\ref{sec:model}), the rms/beam
        in the image, and the total size of the image.}
      \label{tab:data}
    \end{deluxetable*}

    For both MUSIC and Bolocam, the data were collected by scanning the
    CSO in a Lissajous pattern with an rms speed of $240''$~s$^{-1}$
    and an amplitude of $3'$ and $4'$ respectively.
    Nightly observations of Uranus or Neptune were made to determine the
    absolute flux calibration of the data, which is estimated to
    have an rms uncertainty of 5.0\% for Bolocam and 6.0\%, 4.6\%, 5.5\%,
    and 9.4\% in the 147, 213, 281, and 337 GHz MUSIC observing bands
    \citep{Sayers2013_2, Sayers2014}.
    Included in all of the above uncertainties is an overall 3.3\% rms
    uncertainty in the normalization of the planetary brightness model
    \citep{Sayers2012}.
    In the case of Bolocam, hourly observations of bright compact objects
    were made in order to obtain a pointing accuracy of $\simeq 5''$ rms
    \citep{Sayers2011}.
    For MUSIC, a single pointing offset derived from the flux calibration
    observations is sufficient to obtain a pointing accuracy of 
    $\simeq 3''$ rms, and therefore no dedicated observations were made
    for the purpose of pointing calibration \citep{Sayers2014}.

    The Bolocam data have been used in several SZ effect analyses, and the
    details of their reduction are described in \citet{Sayers2011},
    \citet{Sayers2013_2}, and
    \citet{Czakon2015}. Briefly, a template for the atmospheric brightness 
    fluctuations is formed from the average bolometer signal at each
    time sample. This template is then correlated and regressed from
    each bolometer's time-ordered data (TOD). Next, a high-pass filter
    with a characteristic frequency of 250~mHz is applied to the TOD.
    Both of these processes attenuate SZ effect signal, and an image-space
    transfer function is computed by reverse-mapping and processing
    a cluster model through the same reduction pipeline. See Table~\ref{tab:data}.

    In order to reduce the MUSIC data, the first step is to locate discontinuous
    jumps in the TOD that appear to be caused by the readout electronics.
    When a discontinuity is found, the TOD for that MKID is flagged and
    removed from the entire 10-minute long observation. In total, approximately
    10\% of the data are flagged.
    The MUSIC instrument moves along with the CSO telescope during the Lissajous scans,
    and this motion relative to the Earth's magnetic field produces
    a response in the MKIDs at the level of $\simeq 10$~mJy peak-to-peak.
    To remove this response, templates of the AZ and EL motion of the telescope are
    correlated and regressed from the resonant-frequency-response TOD of each
    MKID (changing magnetic fields do not produce a signal in the 
    dissipation-response TOD).\footnote{
      The MUSIC readout works by using a single probe tone centered on the 
      nominal resonant frequency of each MKID
      \citep{Duan2010}. Changes in the resonance depth and/or
      frequency thus produce a change in the complex-valued transmission
      of this probe tone.
      A full calibration sweep of the resonance is performed once every 30 minutes,
      and this sweep provides a mapping from the complex-valued TOD to
      separate real-valued TODs that represent
      the change in the resonance frequency and resonance depth (or dissipation).
      For MUSIC, the signal-to-noise for astronomical sources is many times
      larger in the frequency response than in the dissipation response,
      and so the dissipation-response TODs are not used in this analysis.}
    Additional noise fluctuations, with $1/f$ and $1/f^2$ power
    spectra, are sourced by gain and phase fluctuations in the MUSIC readout
    electronics.
    Templates for these fluctuations are constructed
    from electronics fluctuation monitors and then subtracted from the MKID TODs.
    Analogous to the Bolocam analysis procedure, a template of the atmospheric brightness fluctuations
    is then subtracted, with a separate template constructed for each of
    the four observing bands.
    A high-pass filter with a characteristic frequency of 500~mHz is 
    then applied to the TOD. Compared to Bolocam, the MUSIC data contain more low-frequency
    noise due to both residual instrumental and atmospheric noise, and therefore
    a more aggressive high-pass filter provides an improved signal-to-noise ratio in the MUSIC data.
    Finally, the overall image-space transfer function of the MUSIC processing
    is determined separately for each observing band using the same procedure
    as for Bolocam. Due to MUSIC's slightly smaller FOV, along with the higher
    frequency high-pass filter applied to the MUSIC TOD, the Bolocam images retain fidelity to
    cluster emission on larger angular scales than the MUSIC images.
    See Table~\ref{tab:data}.

  \subsection{Point Source Subtraction}
    \label{sec:point_source}
  
\begin{figure}
      \includegraphics[width=\columnwidth]{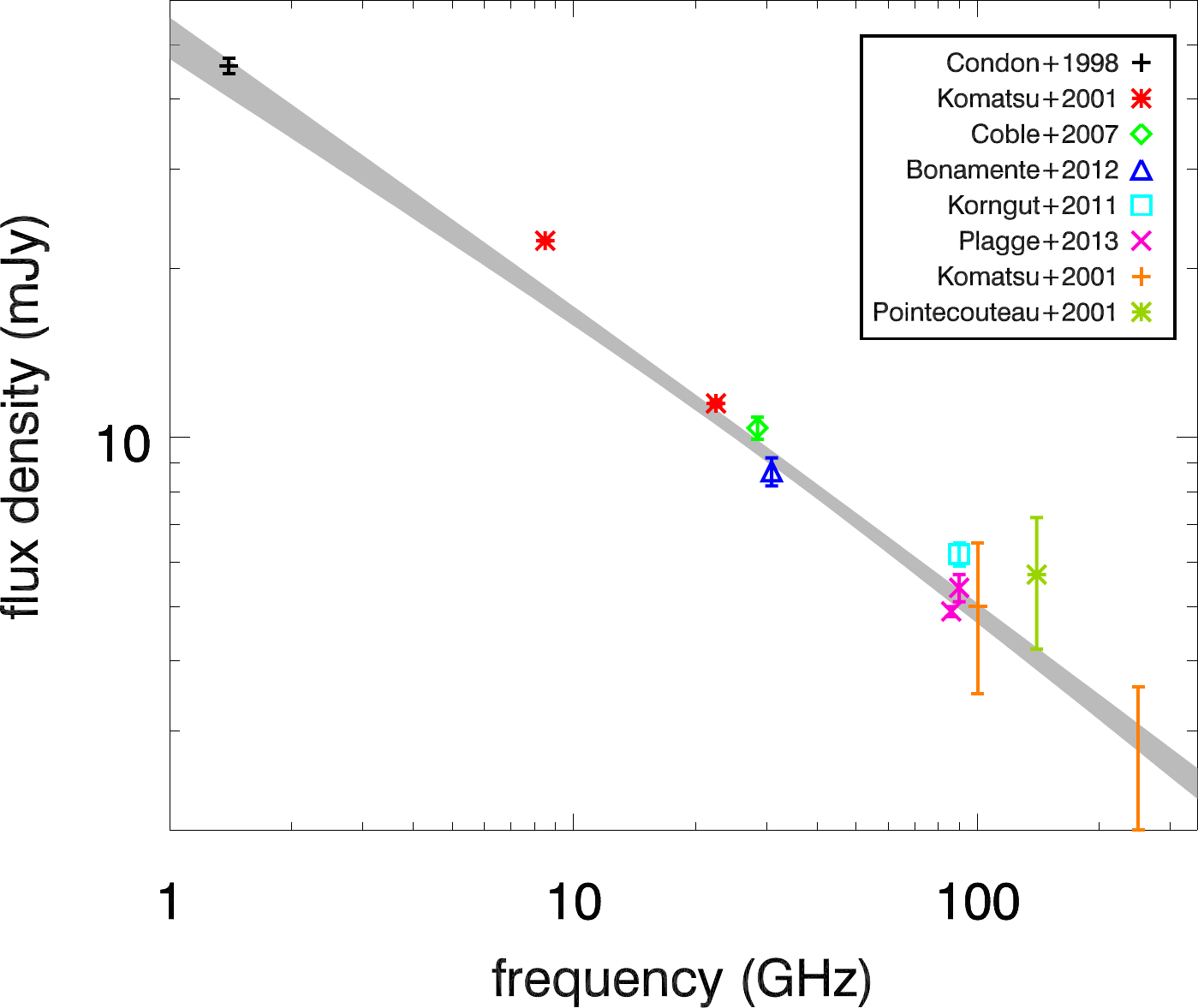}
      \caption{Measurements of the flux density of the AGN in the BCG of \rxj\
        at frequencies close to the \bplusm\ observing bands
        (see text for more details). Overlaid in grey is the 
        68\% confidence region for a power-law fit obtained
        using \textsc{linmix\_err}. This fit has a fractional
        uncertainty of 10\% at the frequencies corresponding
        to the \bplusm\ bands at 140--337~GHz.}
      \label{fig:AGN}
    \end{figure}

    The BCG of \rxj\ hosts a strong AGN, and its emission
    is non-negligible compared to the SZ effect signal in
    the \bplusm\ observing bands. Due to 
    \bplusm's relatively coarse angular
    resolution, and the fact that the AGN is spatially coincident with
    the peak in the SZ effect signal, it is not possible to model and
    subtract the AGN emission using \bplusm\ data alone. As a result,
    a wide range of available measurements of the AGN at frequencies
    close to the \bplusm\ observing bands were used to obtain a power-law 
    fit to the flux density of the AGN.
    Specifically, these data include a 1.4~GHz measurement from
    NVSS \citep{Condon1998}, 8.5 and 22.5~GHz measurements from
    VLA \citep{Komatsu2001}, a 28.5~GHz measurement from OVRO/BIMA
    \citep{Coble2007}, a 30.9~GHz measurement from SZA \citep{Bonamente2012},
    a 90~GHz measurement from MUSTANG \citep{Korngut2011}, 
    86 and 90~GHz measurements from CARMA \citep{Plagge2013}, 
    a 100~GHz measurement from NMA \citep{Komatsu1999}, 
    a 140~GHz measurement from Diablo \citep{Pointecouteau2001},
    and a 250~GHz measurement from SCUBA \citep{Komatsu1999}.
    All of these data were collected using interferometers except for
    MUSTANG (which is a high angular-resolution single-dish photometer), 
    Diablo (where the SZ effect and AGN signals were simultaneously modeled),
    and SCUBA (which was relatively close to the null in the thermal SZ
    effect signal).
    See Figure~\ref{fig:AGN}.
    The best fit obtained using \textsc{linmix\_err} \citep{Kelly2007}, which accounts
    for intrinsic scatter between measurements due to potential, unaccounted-for instrumental
    systematics and/or source variability, is
    \begin{displaymath}
      F(\nu) = (59.1\ \textrm{mJy}) \times \left( \frac{\nu}{\textrm{GHz}} \right)^{-0.527}.
    \end{displaymath}
    Extrapolating this fit to the \bplusm\ observing bands yields
    flux densities of 4.37, 4.26, 3.50, 3.03, and 2.75~mJy at 
    140, 147, 213, 281, and 337~GHz.
    Rounded to the nearest percent,
    the rms uncertainty on the extrapolation is 10\% in 
    all of the \bplusm\ observing bands, and this uncertainty 
    is included in the noise estimates as described in
    Section~\ref{sec:noise} below.

    In addition to the BCG, the \bplusm\ images also contain a non-negligible
    signal from the population of dusty star-forming galaxies behind
    \rxj, although none of these galaxies are individually detected at high significance
    in the \bplusm\ data.
    The emission from these sources has been subtracted according to
    the general procedure described in detail in the Appendix of \citet{Sayers2013_3}
    with some slight modifications described below.
    Briefly, a list of candidate galaxies is compiled from the
    three-band \spire\ observations at 600, 850, and 1200~GHz.
    In total, 192 galaxy candidates are detected by \spire\
    with a signal-to-noise above 3 in at least one of the \spire\ bands.
    For each galaxy candidate, a greybody spectrum is fit to the \spire\
    measurements and is extrapolated to the \bplusm\ observing
    bands.
    For each band, the source positions and brightness estimates for all of the galaxies are combined
    into a single model of the background galaxy emission.
    Next, these models are fit to the \bplusm\ data, floating the value of the overall
    normalization separately for each observing band (see Table~\ref{tab:spire}).
    The average normalization over the five bands is $1.36 \pm 0.23$, indicating that the 
    background emission from the star-forming galaxies is slightly
    brighter than the greybody extrapolation in the \bplusm\ bands.
    Therefore, the \spire\ model is multiplied by a factor of 1.36 prior to subtracting
    it from the \bplusm\ data, and the rms uncertainty on this scaling factor is taken to
    be 0.23 (see Section~\ref{sec:noise}).
    
    Although a single scaling factor for the \spire\ model is
    assumed for all five \bplusm\ observing bands, 
    it is possible that the slight mismatch between the greybody extrapolation
    and the \bplusm\ data is due to the assumed greybody shape rather than the overall
    normalization.
    However, the \bplusm\ measurements are not precise enough to distinguish
    between the two scenarios, and so the simpler case of an overall normalization
    factor is used for this analysis.
    Furthermore, as described in Section~\ref{sec:spectral_fit}, the extreme case
    of not subtracting any model of the dusty star-forming galaxies results in only
    modest changes to the derived ICM constraints, and consequently any minor differences
    between the assumed and true greybody models are unlikely to have a significant
    impact on the overall results.

\begin{figure*}
      \centering \includegraphics[height=0.6\textheight]{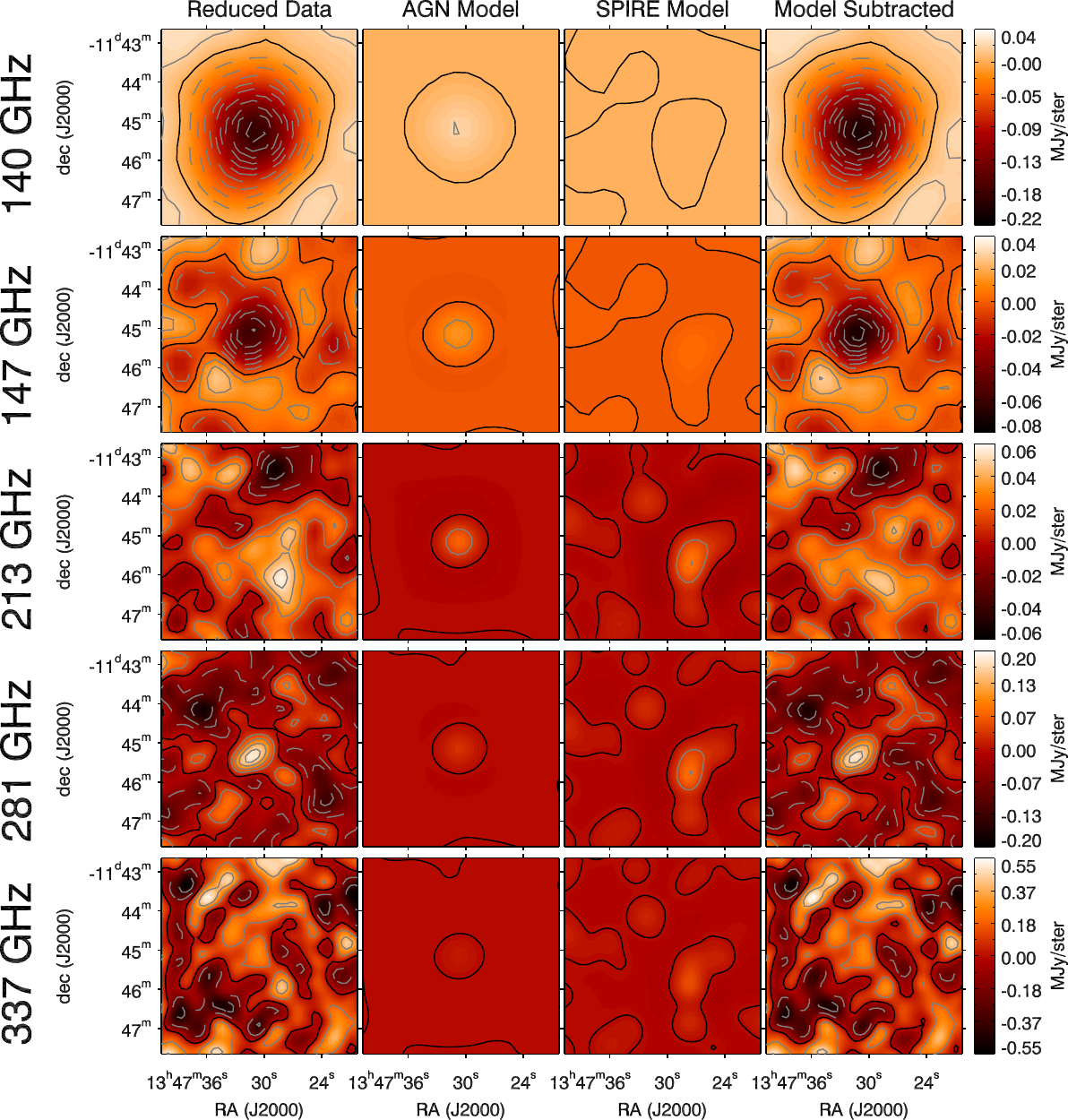}
      \caption{From left to right the columns show thumbnails of the 
        \bplusm\ data, the model of the AGN emission from the cluster BCG, the \spire-derived 
        model for the brightest dusty star-forming galaxies, and the \bplusm\ data
        after subtracting both of these models. For MUSIC, the solid grey lines show positive
        S/N in steps of 1, and the dashed grey lines show negative S/N
        in steps of 1 (for Bolocam the lines
        indicate steps of $\pm 3$). The solid black line is at 0. In all cases, the color
        scale has been stretched to cover the maximum and minimum image
        values. All of the images have been filtered, and the more aggressive
        high-pass filter applied to the MUSIC data results in a more compact
        size for SZ effect signal compared to the Bolocam data.}
      \label{fig:agn_thumbnail}
    \end{figure*}

    Images of the \bplusm\ data, along with the background dusty star-forming galaxy and AGN models, are shown
    in Figure~\ref{fig:agn_thumbnail}.
    The potential impact of these point sources on the SZ effect constraints
    is explored in more detail in Section~\ref{sec:spectral_fit}.

    \begin{deluxetable}{ccccccc} 
      \tablewidth{0pt}
      \tablecaption{\spire\ Dusty Star-Forming Galaxy Model}
      \tablehead{\colhead{band (GHz)} & \colhead{140} & \colhead{147} &
        \colhead{213} & \colhead{281} & \colhead{337} & \colhead{total}} 
      \startdata 
        normalization & $-0.73$ & 2.27 & 1.47 & 1.12 & 2.24 & 1.36 \\
        uncertainty  & \phs1.53 & 1.58 & 0.34 & 0.36 & 0.80 & 0.23 
      \enddata
      \tablecomments{\bplusm-derived fits to the normalization of the dusty star-forming galaxy
        model obtained from \spire. Averaging all four bands together, a best-fit
        normalization of $1.36 \pm 0.23$ is obtained, indicating that the 
        emission in the \bplusm\ observing bands is 36\% brighter than
        the \spire\ greybody extrapolation.}
      \label{tab:spire}
    \end{deluxetable}

  \subsection{Noise Estimation}
   \label{sec:noise}

    The \bplusm\ data contain noise from several different sources, including the 
    instruments themselves, the random arrival of photons, 
    the atmosphere, and the astronomical sky. In all
    cases, the noise is characterized by a set of 1000 separate five-band image 
    realizations which are used to construct a noise covariance
    matrix in a bootstrap fashion (see Section~\ref{sec:model}).
    Some of the noise in each of the 1000 realizations
    is correlated between the separate bands, and therefore a single 
    noise realization contains images for all five bands.
    Each of these noise realizations is constructed as described below, and the
    effects of each noise source on constraining the SZ effect signal are
    detailed in Section~\ref{sec:model}.
    \begin{itemize}
      \item{Instrument, photon, and atmosphere: A jackknife image is formed
        by multiplying the data in a randomly selected subset of half the 
        observations by $-1$ prior to producing the image. This jackknife
        preserves the noise properties of the instrument, random photon, and atmospheric
        brightness fluctuations, including correlations between observing bands,
        but it removes all astronomical signals.
        This is the dominant noise in the data, particularly in the four MUSIC
        bands.}
      \item{AGN: The template image of the AGN is different for each observing band,
        and consists of that band's PSF, after accounting for the filtering effects
        of the data processing, multiplied by the estimated flux density of the AGN
        in that band. Based on the fit presented in Section~\ref{sec:point_source},
        a 10\% rms uncertainty was assigned to the estimated flux 
        density of the AGN.
        Furthermore, because any fractional deviation in the source brightness from the model
        will be common among all five \bplusm\ observing bands, the noise fluctuations
        are assumed to be completely correlated between the five observing bands.
        Specifically, the AGN template in the five bands is multiplied by
        a single, common Gaussian random value drawn from a distribution with a mean of 1.0
        and a standard deviation of 0.1.}
      \item{\spire-detected galaxies: The normalization of the \spire-derived model was
        determined to be $1.36 \pm 0.23$, and a Gaussian random value based
        on this constraint is used to describe the uncertainty on this model.
        For each observing band, a template is constructed based on the \spire\
        model and that band's PSF and data processing filter.
        As with the AGN template, the noise fluctuations are assumed to be completely correlated
        between all five observing bands due to the fact that a fractional change
        in source brightness in one band will result in the same fractional change
        in the other four bands given the nominal greybody fit.}
      \item{Undetected dusty star-forming galaxies: Although \spire\ detects 192 galaxy
        candidates towards \rxj, the bulk of the population of dusty
        star-forming galaxies is below the \spire\ detection limit. In order
        to account for brightness fluctuations due to the non-uniform
        spatial distribution of these galaxies, a random sky 
        realization is generated based on the model of \citet{Bethermin2011}, after
        correcting the model for the galaxies that were detected by 
        \spire\ and subtracted (see the Appendix of \citet{Sayers2013_3} for more
        details). A single sky realization is used for all five observing
        bands in order to preserve the correlations between bands.
        For each band, the sky realization is
        convolved with the PSF and data processing filter.}
      \item{Primary CMB fluctuations: A random sky realization
        is generated based on the power spectrum measurements
        from \citet{Planck2015_XI} and is added to the noise realizations
        for each of the five observing bands after correcting
        for each band's PSF and data-processing filter.
        As with the sky realizations of the undetected dusty star-forming
        galaxies, a single sky realization is used for all five bands
        in order to preserve the correlations between bands.}
      \item{Flux Calibration: The flux calibration uncertainty includes
        two separate components. The overall planetary brightness model
        has an uncertainty of 3.3\%, and fluctuations based on this
        uncertainty are assumed to be completely correlated between
        all five observing bands. In addition, as described in 
        Section~\ref{sec:cal}, there is an uncorrelated measurement
        uncertainty on the flux calibration for 
        each observing band.}
    \end{itemize}

\section{X-ray Data Reduction}
\label{sec:X-ray}

  As described below in Section~\ref{sec:spectral_fit}, a \chandra\ X-ray
  constraint on the ICM temperature \Te\ is included in the overall analysis.
  This measurement of \Te\ was obtained from
  the longest \chandra\ observation of \rxj, ID 14407, 
  which was reprocessed using CIAO version 4.7 (December 2014) and
  CALDB 4.6.9 (September 2015). The observation is very clean,
  with 62.9 ksec from the 63.2 ksec exposure remaining after
  light curve filtering to reject any 200-second bins where
  the count rate varied by more than 3$\sigma$ from the median.
  The standard ACIS background file was scaled to match the
  off-source count rate in the 10--12 keV energy range, and 
  subtracted. Spectral fits were performed with the 0.7--9.5 keV
  band using a photon-absorbed APEC plasma model with 
  n$_{\textrm{H}}$~=~$4.60 \times 10^{20}$~cm$^2$ (the Leiden-Argentine-Bonn value
  for \rxj's sky position). Excising the extremely bright
  cool-core at a radius of 150~kpc ($0.43'$) results in
  an emission-weighted $T_{\mathrm{e}} = 13.4 \pm 0.7$~keV
  within \rfive. 
  This measurement differs somewhat compared to previous X-ray measurements of \Te. 
  For example, \citealt{Mahdavi2013} find $T_{\mathrm{e}} = 12.1 \pm 0.4$~keV
  based on a combined \chandra--{\it XMM} analysis. Furthermore,
  there is substantial evidence that the ICM is not isothermal
  within \rfive\ (e.g., \citealt{Johnson2012}).
  Section~\ref{sec:spectral_fit} explores the impact of these different
  \Te\ values on the overall ICM constraints.

\section{Spatial Model of the SZ Effect Signal}
  \label{sec:model}

  \begin{figure*}
    \centering \includegraphics[height=0.6\textheight]{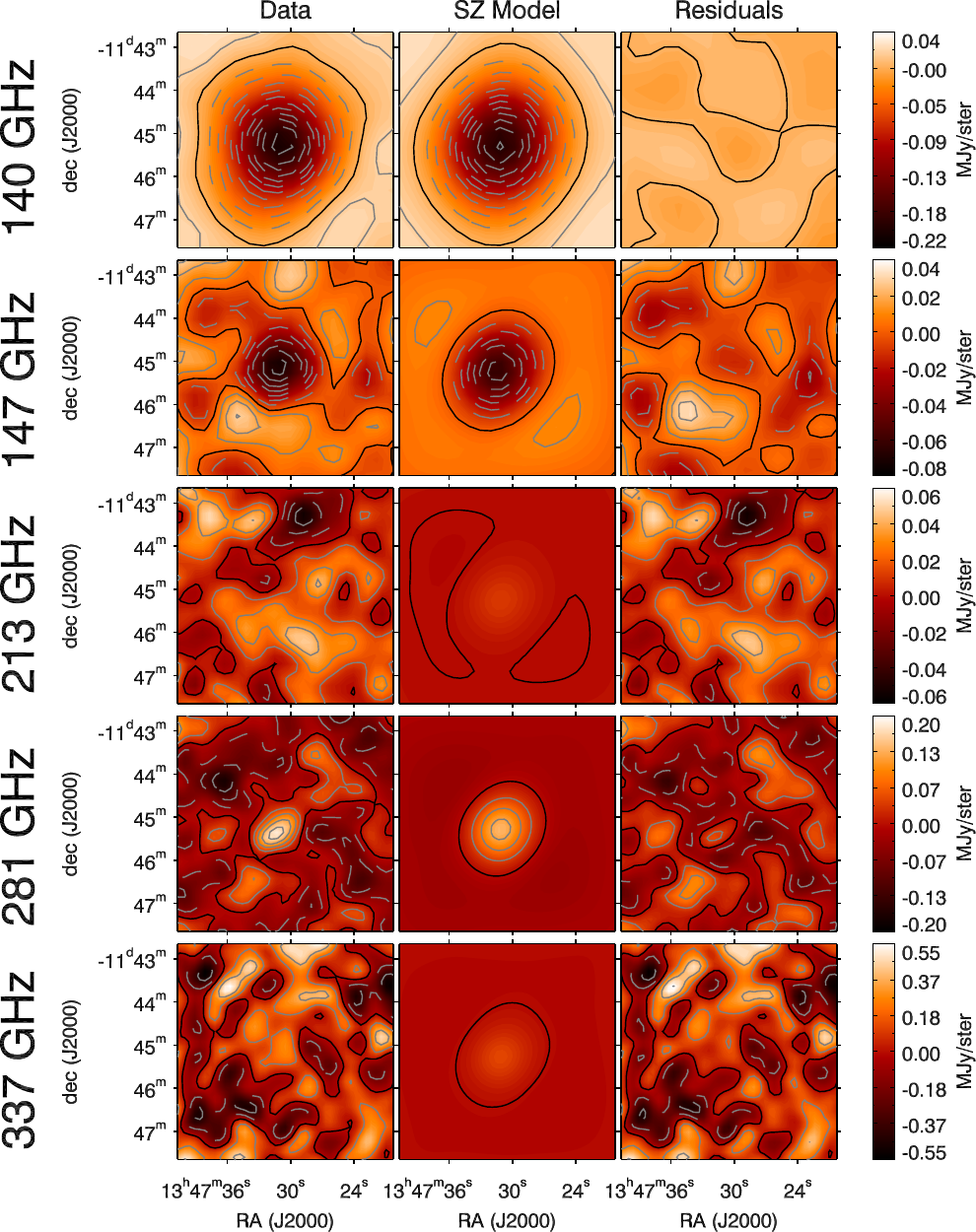}
    \caption{From left to right the columns show thumbnails of the 
      \bplusm\ data after subtracting the AGN and dusty star-forming galaxy model, 
      the best-fit model of the SZ effect signal based on the elliptical
      gNFW model from Table~2 of \citet{Czakon2015}, and the residuals after subtracting the SZ
      effect model.
      In all cases, the SZ effect model provides a good fit to the
      data, and the residuals are consistent with noise.}
    \label{fig:model_thumbnail}
  \end{figure*}

\begin{figure*}
    \centering
    \includegraphics[width = 0.85\textwidth]{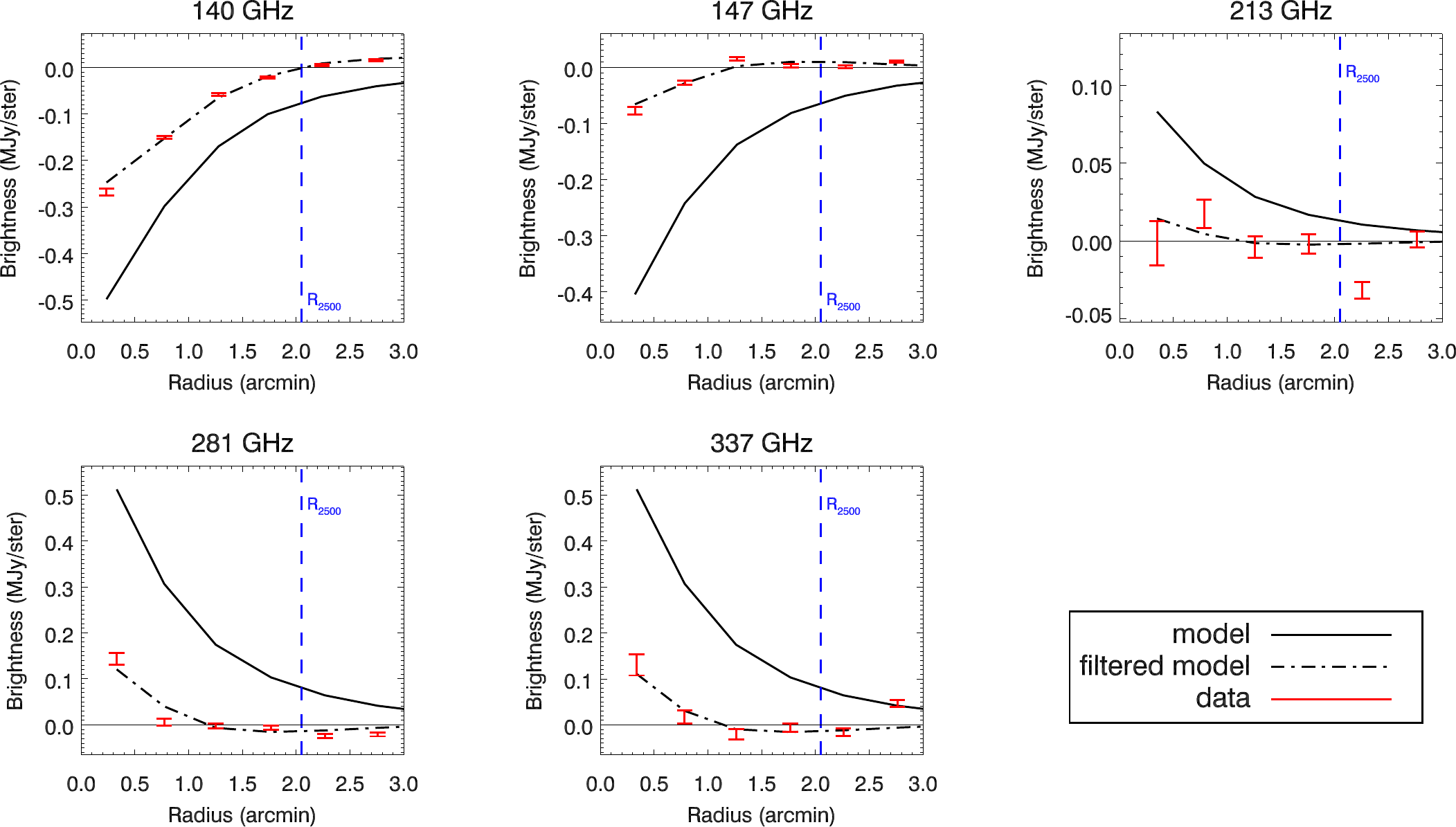}
    \caption{Radial profiles of the SZ data and the elliptical gNFW model
      used to describe it. In order to compare the model to the data, it
      must first be convolved with the transfer function of the data processing
      to determine how it will appear after this filtering. The best-fit model
      to the data is shown in black, which a solid line representing the
      model prior to filtering and the dot-dashed line representing the
      model after filtering. The filtering attenuates signal on large
      angular scales, resulting in a peak with smaller absolute brightness.
      In addition, the filtered signal passes through zero at intermediate
      radii, and then takes on the opposite sign at large radii.
      The data, with error bars, are shown in red and produce a good fit
      quality to the model in all five bands. The location of \rfive\
      is shown as a blue dashed line.}
    \label{fig:profiles}
  \end{figure*}

  Although \rxj\ shows evidence for a shock in the ICM to the SE of the
  cluster core \citep{Mason2010, Johnson2012, Kreisch2014}, its X-ray morphology indicates
  that it is among the most relaxed known clusters \citep{Mantz2015}.
  As a result, the best-fit elliptical generalized Navarro, Frenk, and White
  (gNFW, \citealt{Nagai2007}) profile from Table~2 of \citet{Czakon2015},
  based on a fit to the Bolocam data, was 
  selected to model the spatial shape of the SZ effect signal.
  This spatial template was then convolved with the PSF and 
  data-processing filter specific to each of the five observing bands.
  The normalization of the template was varied as a free 
  parameter separately for each observing band using all of the
  data within the overdensity
  radius \rfive\ = 0.71~Mpc ($2.05'$) 
  published in \citet{Czakon2015} based on the techniques
  described in \citet{Mantz2010}.\footnote{
    \rfive\ is the radius enclosing an average density 2500 times the
    critical density of the universe. As shown in Table~\ref{tab:data},
    the \bplusm\ data have good sensitivity to angular scales
    corresponding to \rfive, allowing for accurate constraints
    within that aperture size.}
  This model produces a good fit quality in all five observing bands,
  as quantified in Table~\ref{tab:chisq} and shown in the thumbnails
  in Figure~\ref{fig:model_thumbnail} and the radial profiles in
  Figure~\ref{fig:profiles}.
  Based on these model fits, the average SZ effect brightness within \rfive\
  was computed for each band, with the results given in
  Table~\ref{tab:brightness}.

  \begin{deluxetable}{ccccc} 
    \tablewidth{0pt}
     \tablecaption{gNFW Model Fit Parameters}
     \tablehead{\colhead{instrument} & \colhead{band} &
       \colhead{$p_0$} & \colhead{$\chi^2$/DOF} &
       \colhead{PTE} \\
       \colhead{} & \colhead{(GHz)} & \colhead{($10^{-11}$ erg cm$^{-3}$)} &
       \colhead{} & \colhead{}} 
     \startdata
       Bolocam & 140 & \phs$34.3 \pm $\phn$1.5$ & 147/118 & 0.04 \\
       MUSIC & 147 & \phs$28.6 \pm $\phn$4.0$ & 239/206 & 0.06 \\
       MUSIC & 213 & $-16.6 \pm $$26.9$ & 456/466 & 0.62 \\
       MUSIC & 281 & \phs$38.6 \pm $$11.2$ & 740/730 & 0.39 \\
       MUSIC & 337 & \phs$24.8 \pm $$26.4$ & 919/971 & 0.88
     \enddata
     \tablecomments{Summary of the gNFW model fits to MUSIC and Bolocam. The shape of the 
     model is based on the fit published in Table~2 of \citet{Czakon2015},
     and the only free parameter in the fits is the normalization
     of the model $p_0$. In the limit of a purely thermal SZ signal, the value of 
     $p_0$ will be the same for all five observing bands.
     Except for the 213~GHz band, which is near the thermal
     SZ effect null and easily contaminated by the kinetic SZ effect signal,
     there is good agreement between the MUSIC and Bolocam measurements. In
     addition, the quality of fit in all five bands is good, indicating
     that the data are consistent with the assumed model.
     The PTE values are based on the theoretical $\chi^2$ distribution,
     and therefore may be slightly biased (see Tables 2 and 3 in \citealt{Sayers2011}).}
     \label{tab:chisq}
  \end{deluxetable}

\begin{figure*}
    \centering
    \includegraphics[width=.75\textwidth]{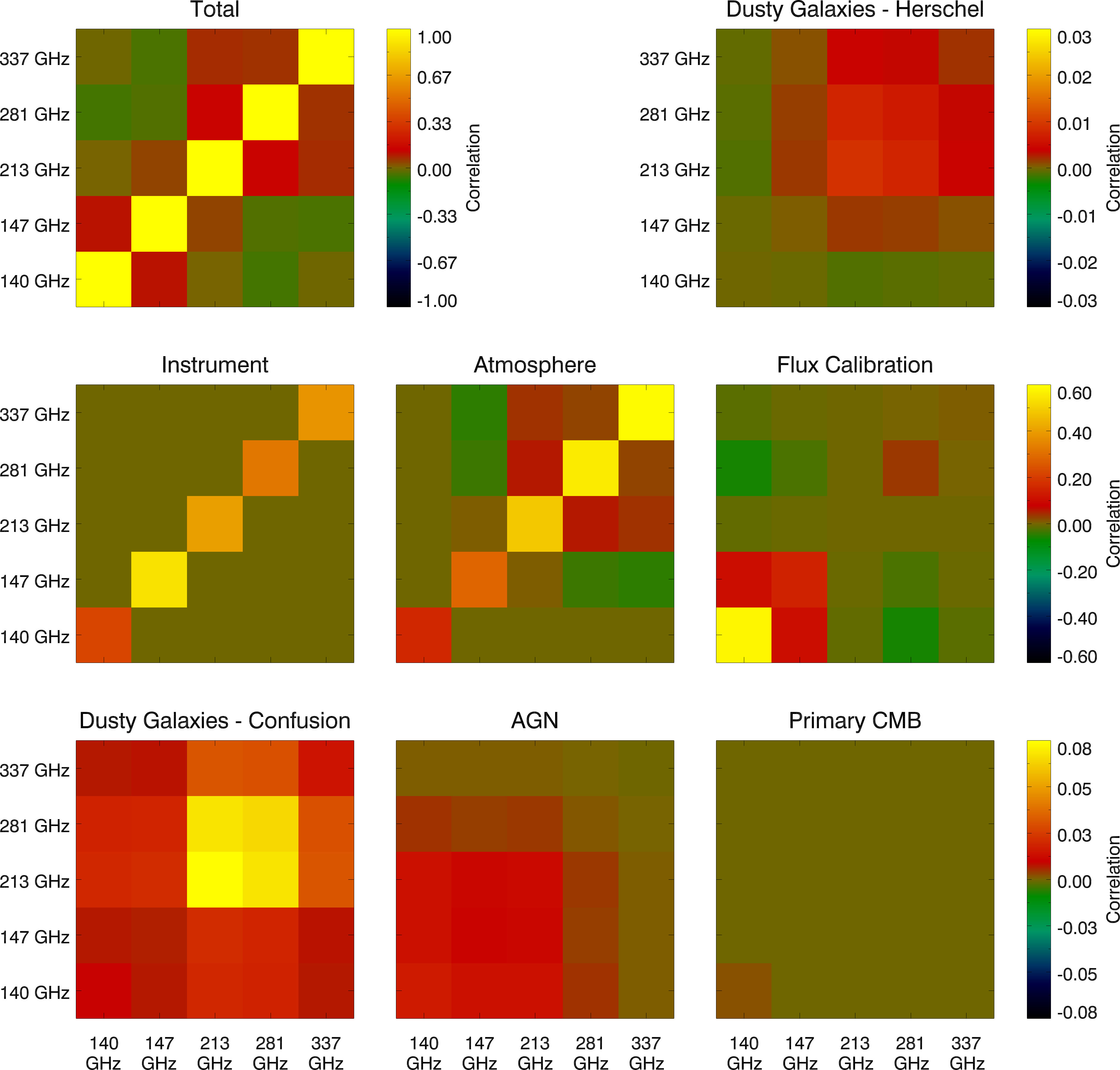}
    \caption{The total covariance matrix is the sum of several
      noise terms, as described in Section~\ref{sec:noise}.
      To illustrate their relative contribution to the total noise,
      the plots above show the covariance matrix of each noise term
      normalized by the diagonal elements of the total covariance
      matrix. Specifically, $C^{\mathrm{tot}} = \sum C^x$ where $C^{\mathrm{tot}}$
      is the total covariance matrix and $C^x$ is the covariance matrix
      of noise term $x$. The plots above show $C^x_{i,j} / 
      \sqrt{C^{\mathrm{tot}}_{i,i} C^{\mathrm{tot}}_{j,j}}$, 
      with the total noise covariance 
      (top left), the \spire\ dusty star-forming galaxy template (top right),
      noise in the jackknife realizations that is not correlated between
      image pixels (middle left, approximately equal to the raw photon and instrumental noise),
      the additional noise in the jackknife realizations that is correlated between
      image pixels (middle, approximately equal to the residual atmospheric noise), 
      flux calibration uncertainties (middle right), 
      dusty star-forming galaxies below the \spire\
      detection limit (bottom left), uncertainties in the model for
      the emission from the central AGN (bottom middle), and primary CMB
      fluctuations (bottom right). Note the different color scales
      for the three rows. For MUSIC, the effective high-pass filtering of 
      the data processing has a characteristic scale of $\ell_0 \simeq 8000$, and
      this filter eliminates nearly all of the primary CMB fluctuations.}
    \label{fig:covariance}
  \end{figure*}

  In performing the model fits, the image pixel weighting and overall
  $\chi^2$ values were computed based on the assumption that the 
  noise is uncorrelated between pixels. Specifically, the 
  image-space pixel sensitivity was computed for each observing band
  based on the standard deviation of the pixel flux density weighted
  by the square root of integration time in the set of 
  1000 noise realizations described in detail in Section~\ref{sec:noise}.
  This assumption is not strictly correct, as many of the noise terms
  are correlated between image pixels and between observing bands.
  Therefore, to fully account for the noise, the spatial template
  of the SZ effect signal was also fit to each of the 1000 noise
  realizations and these fits were used to compute the full
  $5 \times 5$ covariance matrix for the SZ effect 
  brightness. The covariance matrix
  is given in Table~\ref{tab:brightness} and the relative contributions
  of the noise terms described in Section~\ref{sec:noise} to this covariance matrix
  are plotted in Figure~\ref{fig:covariance}. Many of the 
  off-diagonal elements of the SZ effect brightness covariance
  matrix are comparable in magnitude to the on-diagonal elements, 
  particularly for adjacent observing
  bands. In the case of MUSIC, which obtains simultaneous multi-band
  images, some of this correlation is due
  to fluctuations in the atmospheric brightness.
  However, the bulk of the band-to-band correlations are due to 
  the unwanted astrophysical signals and the common planetary
  calibration model.
  In addition, primary CMB fluctuations are not
  a significant noise term in any of the four MUSIC observing bands
  due to the high-pass filter applied when processing the data,
  which corresponds to a characteristic angular multipole of 
  $\ell_0 \simeq 7000$--9000 (see Table~\ref{tab:data}).

  \begin{deluxetable}{crrrrr} 
    \tablewidth{0pt}
     \tablecaption{SZ Effect Brightness}
     \tablehead{\colhead{band (GHz)} & \colhead{140} &
       \colhead{147} & \colhead{213} &
       \colhead{281} & \colhead{337}} 
     \startdata
       \mjy $\times 10^{-2}$ & $-17.7$ & $-14.3$ & 2.9 & 18.2 & 18.2 \\ 
       \cutinhead{Covariance [\mjy]$^2 \times 10^{-4}$} 
       140 GHz & 1.3  & 0.3  & 0.1  & $-0.3$   & 0.0 \\
       147 GHz & ---          & 4.8  & 0.4  & $-0.3$   & $-1.5$ \\
       213 GHz & ---          & ---          & 21.7     & 3.6  & 7.3 \\
       281 GHz & ---          & ---          & ---      & 27.9     & 7.0 \\
       337 GHz & ---          & ---          & ---      & ---          & 374.4
     \enddata
     \tablecomments{The average surface brightness within \rfive\ derived
       from the SZ effect model fits. The top row gives the best fit
       surface brightness and the bottom set of $5 \times 5$ entries
       gives the covariance matrix derived from fits to the 1000
       noise realizations.}
     \label{tab:brightness}
  \end{deluxetable}

  A range of other models from the literature were also fit to the 
  data in order to explore how the SZ effect brightness might change
  as a result of the chosen spatial template of the SZ effect signal.
  These other models included the average profiles from \citet{Arnaud2010}, 
  \citet{Planck2013_V}, and \citet{Sayers2013_2}
  along with the \rxj\ specific profile from \citet{Bonamente2012}.
  None of these models produced an acceptable fit quality to the
  140~GHz Bolocam data, with probabilities to exceed (PTEs) below 0.001 in all cases.\footnote{
    The poor fit quality of the ensemble-average profiles
    from \citet{Arnaud2010}, \citet{Planck2013_V}, and \citet{Sayers2013_2}
    to the Bolocam data
    is not surprising, given the large amount of cluster-to-cluster
    scatter relative to those average profiles.
    In contrast, and analogous to the nominal \citet{Czakon2015} profile used in this analysis,
    the \citet{Bonamente2012} profile is based on a fit specifically to \rxj.
    However, the \citet{Bonamente2012} profile is more centrally peaked than the
    \citet{Czakon2015} profile, and this difference is the most likely cause of 
    the poor fit quality of the \citet{Bonamente2012} profile to the Bolocam data.
    The reason for the difference between the profiles derived by
    \citet{Czakon2015} and \citet{Bonamente2012}
    is unclear, although several possibilities exist, such as 
    differences in measurement technique, observing frequency,
    and point-source treatment.}
  As a result, none of these models are suitable for exploring the model
  dependence of the results.

\newpage
\section{SZ Effect Spectral Fits}
  \label{sec:spectral_fit}

  The total SZ effect brightness can be described according to 
  \begin{eqnarray*}
    I(\nu) & = & I(\nu)_{\mathrm{tSZ}} + I(\nu)_{\mathrm{kSZ}} \\
    I(\nu)_{\mathrm{tSZ}} & = & f(\nu,T_{\mathrm{e}}) h(\nu) y \\
    I(\nu)_{\mathrm{kSZ}} & = & 
       -R_{\mathrm{kSZ}}(\nu,T_{\mathrm{e}}) h(\nu) \frac{v_{\mathrm{pec}}}{c} \tau_{\mathrm{e}} \\
    y      & = & \int \frac{k_B \sigma_T}{m_{\mathrm{e}} c^2} n_{\mathrm{e}} T_{\mathrm{e}} dl \\
    \tau_{\mathrm{e}} & = & \int \sigma_T n_{\mathrm{e}} dl
  \end{eqnarray*}
  where $\nu$ is the observing frequency, 
  $k_B$ is Boltzmann's constant, $\sigma_T$ is the Thompson
  cross section, $m_{\mathrm{e}}$ is the electron mass, $c$ is the speed
  of light, $n_{\mathrm{e}}$ and \Te\ are the electron density and temperature,
  \vpec\ is the ICM line of sight peculiar velocity, 
  $f(\nu,T_{\mathrm{e}})$ describes the spectral dependence of the thermal
  SZ effect signal including relativistic corrections, $h(\nu)$ 
  describes the conversion from a CMB temperature fluctuation
  to a brightness fluctuation in \mjy, $R_{\mathrm{kSZ}}(\nu, T_{\mathrm{e}})$
  describes the relativistic corrections to the kinetic SZ effect
  signal, and $dl$ is the differential element along the line
  of sight.
  Based on these equations, the total SZ effect signal is dictated by four physical properties
  of the ICM; \Te, $y$, \taue, and \vpec.
  Although the five-band SZ effect brightness measurements described in 
  Section~\ref{sec:model} could
  in principle constrain all four of these ICM properties,
  a prior for the value of \Te\ is employed based on the
  X-ray measurements from \chandra\ described in 
  Section~\ref{sec:X-ray}. 
  In addition, the ICM is assumed to be isothermal, which implies that 
  $y \propto T_{\mathrm{e}} \tau_{\mathrm{e}}$.

\begin{figure*}
    \centering
    \includegraphics[height=0.26\textwidth]{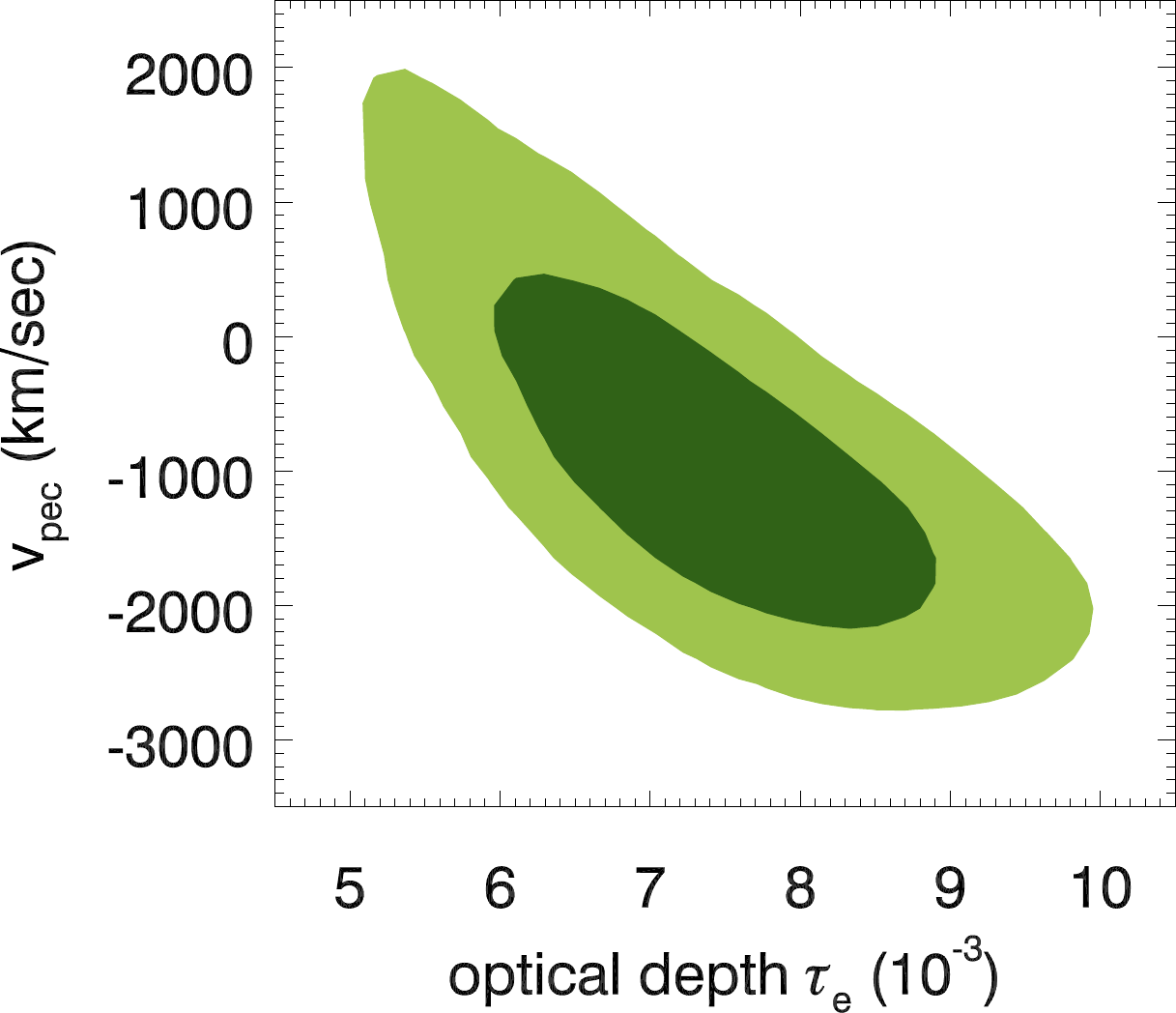}
    \hspace{0.02\textwidth}
    \includegraphics[height=0.26\textwidth]{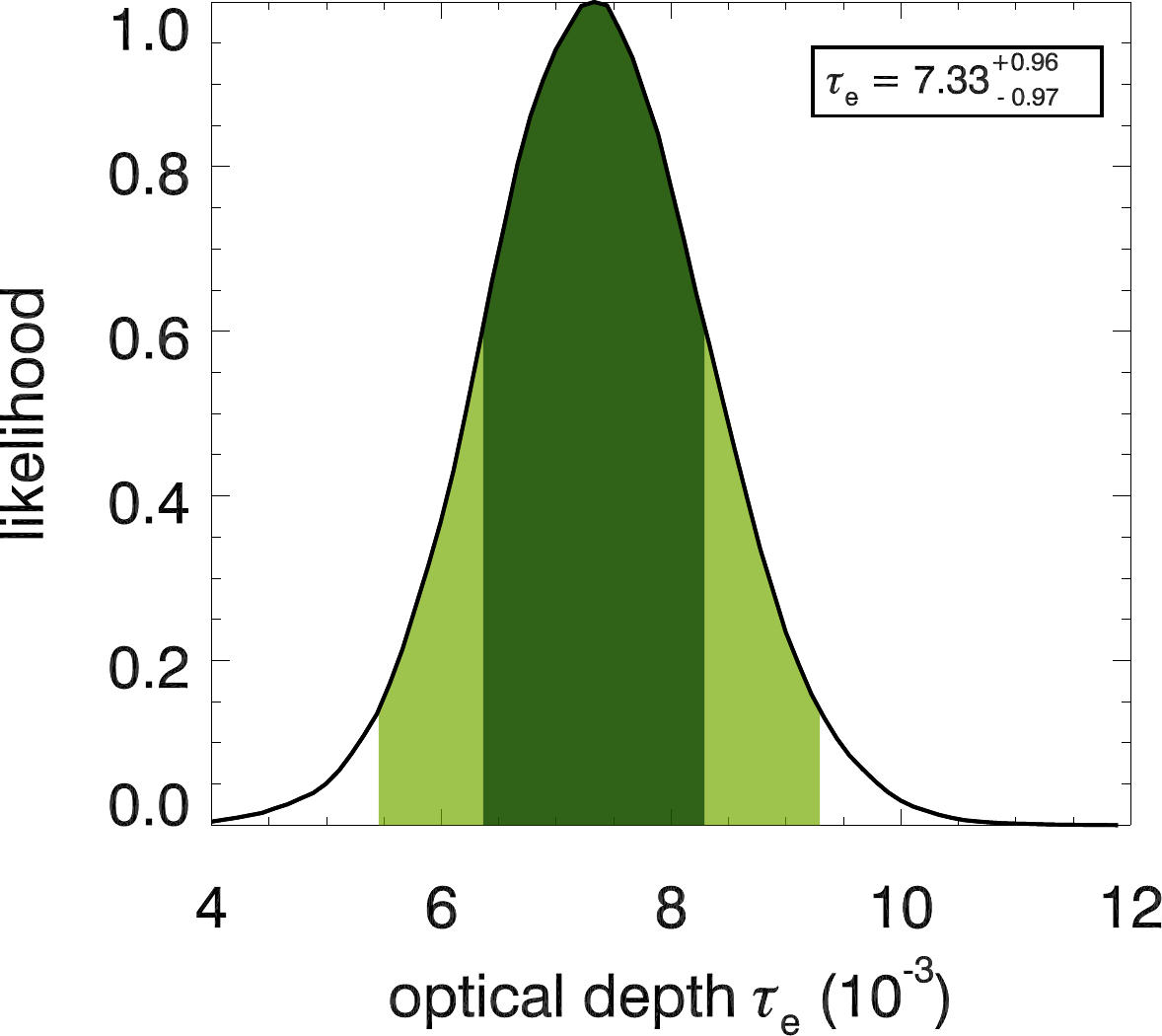}
    \hspace{0.02\textwidth}
    \includegraphics[height=0.26\textwidth]{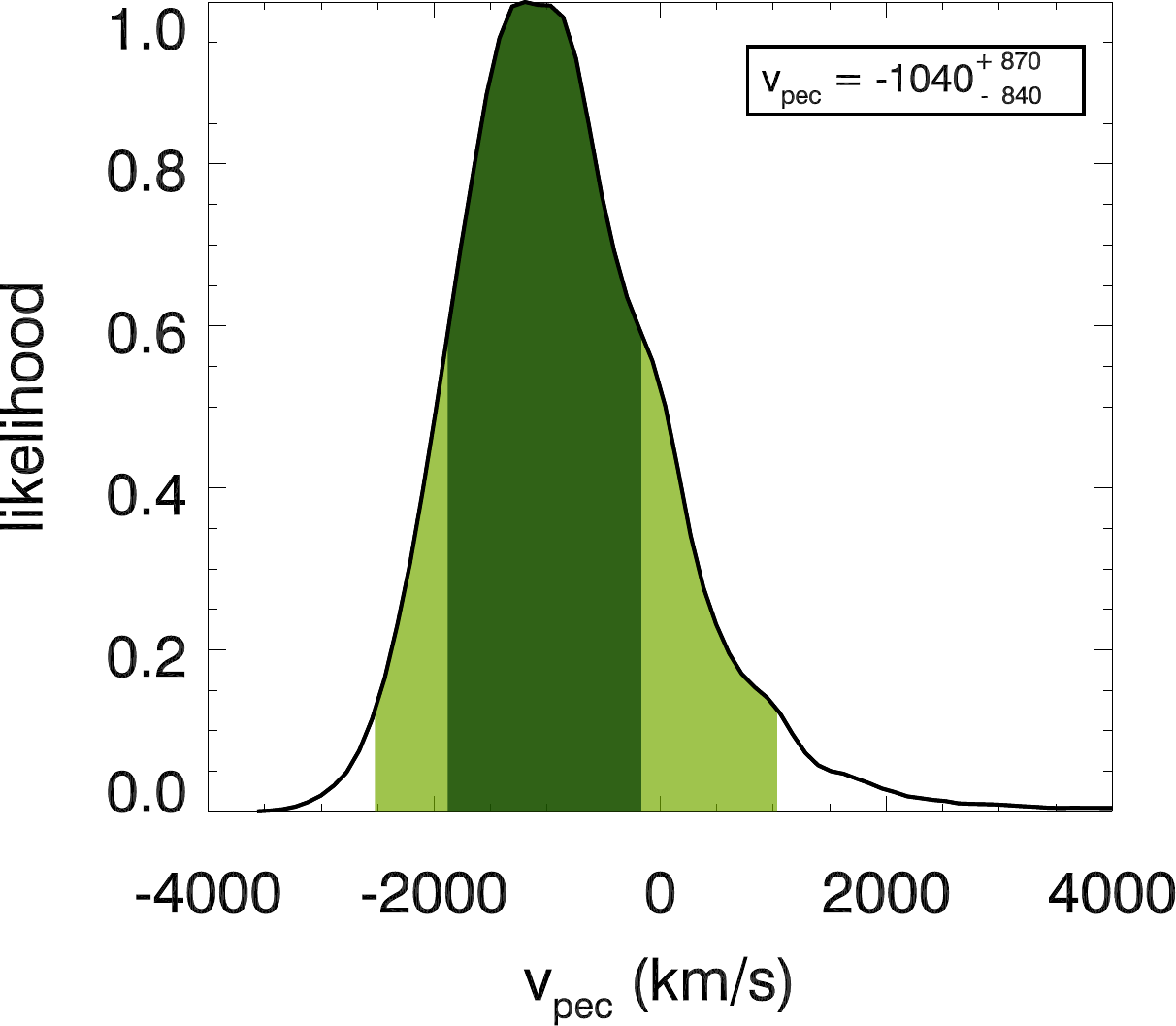}
    \caption{Posterior distributions from the MCMC for the optical depth \taue\ and line
      of sight velocity \vpec\ derived from the five-band SZ
      effect brightness measurements. Light green indicates
      the 95\% confidence region and dark green indicates the
      68\% confidence region. The marginalized 68\% confidence
      regions are $\tau_{\mathrm{e}} = 7.33^{+0.96}_{-0.97} \times 10^{-3}$
      and $v_{\mathrm{pec}} = -1040^{+870}_{-840}$~km/s.}
    \label{fig:likelihood}
  \end{figure*}

\begin{figure}
    \includegraphics[width=\columnwidth]{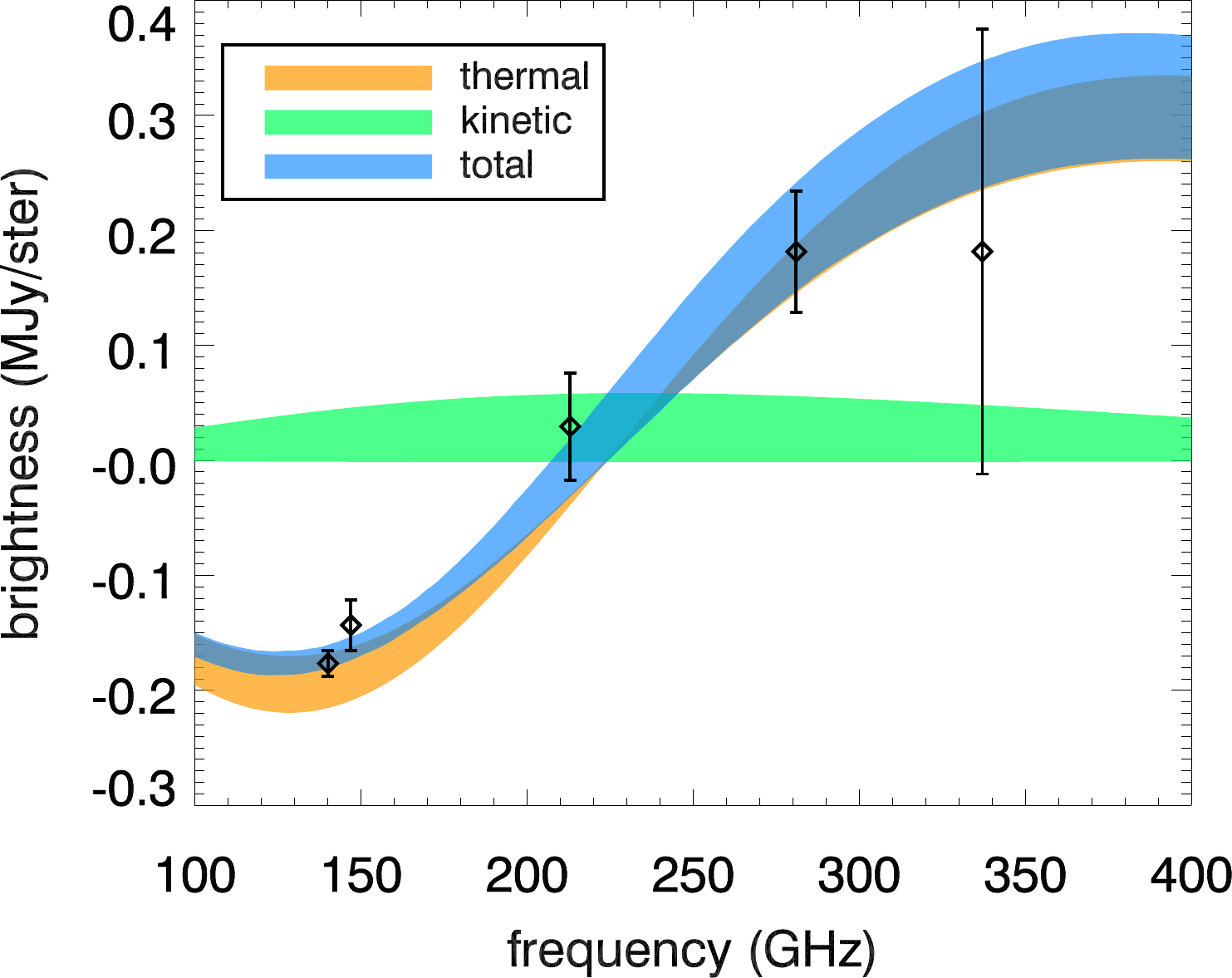}
    \caption{Five-band SZ effect brightness measurements
      within \rfive\ towards \rxj. The error bars denote the
      square root of the diagonal elements of the covariance
      matrix. Overlaid in orange, green, and blue are the 68\%
      confidence regions for the thermal, kinetic, and total
      SZ effect brightness.}
    \label{fig:sz_spectrum}
  \end{figure}

  A Markov chain Monte Carlo (MCMC) is employed to sample the parameter
  space in \taue\ and \vpec\ allowed by the five-band SZ effect
  brightness measurements given in Table~\ref{tab:brightness} along
  with the X-ray prior on \Te. For each step in the chain,
  the SZ effect spectrum is computed using the \textsc{SZpack} code
  described in \citet{Chluba2012, Chluba2013} with the `3D'
  setting.
  The results of the MCMC are presented in Figure~\ref{fig:likelihood},
  which shows the clear degeneracy between the values of \taue\ and \vpec.
  Separately marginalizing over each parameter yields 68\% confidence
  regions of $\tau_{\mathrm{e}} = 7.33^{+0.96}_{-0.97} \times 10^{-3}$
  and $v_{\mathrm{pec}} = -1040^{+870}_{-840}$~km~s$^{-1}$.
  A plot of the 68\% confidence regions for the thermal, kinetic, and
  total SZ effect brightness is shown in Figure~\ref{fig:sz_spectrum},
  with the five-band brightness measurements overlaid.

  In order to test the consistency of the SZ effect brightness measurements
  in the five separate observing bands, along with providing an understanding
  of the contribution of each observing band to the overall constraints
  on \taue\ and \vpec, five additional fits were performed after
  discarding the data from one of the observing bands.
  In all cases, the best-fit values of \taue\ and \vpec\ shift by less
  than the 68\% uncertainties, and in most cases the shift is much less
  than the uncertainty value.
  In addition, the uncertainties on \taue\ and \vpec\ generally do not
  increase significantly when one band is dropped, although dropping the 
  281~GHz data results in uncertainties that are approximately 50\% larger
  for both \taue\ and \vpec. This degradation
  occurs because the 140--213~GHz measurements do not provide
  a sufficient spectral range to fully disentangle the thermal and kinetic
  SZ effect signals, and the 337~GHz data exhibit significant noise and do
  not provide a strong constraint.
  In addition, removing the 140~GHz Bolocam data, which 
  are approximately two times more sensitive to the SZ effect 
  than the 147~GHz MUSIC data,
  only results in a $\simeq 15$\% degradation to the \vpec\ constraint.
  This is due to the fact that the separation of the thermal and kinetic
  SZ effect signals 
  is limited almost entirely by the uncertainty on the 281~GHz measurement. 

  \begin{deluxetable}{ccccccc} 
    \tablewidth{0pt}
     \tablecaption{ICM Optical Depth and Velocity Constraints}
     \tablehead{\colhead{140} & \colhead{147} & \colhead{213} &
       \colhead{281} & \colhead{337} & \colhead{\taue} & \colhead{\vpec} \\
       \colhead{GHz} & \colhead{GHz} & \colhead{GHz} &
       \colhead{GHz} & \colhead{GHz} & \colhead{$10^{-3}$} & \colhead{km s$^{-1}$}}
     \startdata
       \vspace{3pt}
       $\checkmark$ & $\checkmark$ & $\checkmark$ & $\checkmark$ & $\checkmark$ & 
         $7.33^{+0.96}_{-0.97}$ & $-1040^{+\phn870}_{-\phn840}$ \\
       \vspace{3pt}
                    & $\checkmark$ & $\checkmark$ & $\checkmark$ & $\checkmark$ & 
         $6.41^{+1.12}_{-1.13}$ & $-1300^{+1110}_{-1080}$ \\
       \vspace{3pt}
       $\checkmark$ &              & $\checkmark$ & $\checkmark$ & $\checkmark$ & 
         $7.34^{+0.95}_{-0.95}$ & \phn$-890^{+\phn890}_{-\phn870}$ \\
       \vspace{3pt}
       $\checkmark$ & $\checkmark$ &              & $\checkmark$ & $\checkmark$ & 
         $7.14^{+0.98}_{-0.97}$ & \phn$-870^{+\phn960}_{-\phn920}$ \\
       \vspace{3pt}
       $\checkmark$ & $\checkmark$ & $\checkmark$ &              & $\checkmark$ & 
         $7.87^{+1.64}_{-1.57}$ & $-1730^{+1440}_{-1180}$ \\
       $\checkmark$ & $\checkmark$ & $\checkmark$ & $\checkmark$ &              & 
         $7.40^{+1.04}_{-1.05}$ & $-1130^{+\phn940}_{-\phn930}$
     \enddata
     \tablecomments{Marginalized 68\% confidence regions for \taue\ and \vpec\
       based on fits to the full data set (top row) and to subsets of the 
       data. The check marks indicate which observing bands were included in the
       fit. Removing the data from one observing band does not cause
       the best-fit values to move outside of the 68\% confidence regions,
       although the uncertainties become significantly larger when 
       the 281~GHz band is excluded. Due to the relatively
       high noise in the 337~GHz band, removing the 281~GHz band significantly
       reduces the spectral coverage of the data and makes it more difficult
       to separate the thermal and kinetic SZ effect signals.}
     \label{tab:vpec}
  \end{deluxetable}

  To estimate the systematic uncertainty associated with the X-ray
  derived value of \Te\ for \rxj, fits were also performed based
  on different X-ray analyses.
  First, if the core region is not excised from the \chandra\
  data, then the best-fit value of \Te\ is $12.7 \pm 0.3$~keV
  (compared to $13.4 \pm 0.7$~keV when the core is excised).
  This value of \Te\ results in $\tau_{\mathrm{e}} = 7.52^{+1.04}_{-1.06} \times 10^{-3}$
  and $v_{\textrm{pec}} = -740^{+960}_{-970}$~\kms, which
  differ by $0.19 \times 10^{-3}$ and 300~\kms\ from the
  measurements based on a core-excised \Te.
  An additional fit was also performed using the value of 
  $T_{\mathrm{e}} = 12.1 \pm 0.4$~keV from
  \citet{Mahdavi2013}, which is based on a joint \chandra/{\it XMM}
  analysis.
  The best fit values of \taue\ and \vpec\ using the
  \citet{Mahdavi2013} measurement
  are $7.89^{+1.01}_{-1.02} \times 10^{-3}$ and $-770^{+800}_{-800}$~\kms, which differ by
  $0.56 \times 10^{-3}$ and 270~\kms\ from the values obtained
  using the nominal measurement of \Te.
  Compared to the measurement uncertainties, the differences due to the X-ray analysis
  are smaller by a factor of approximately two for \taue\ and by a factor of 
  approximately three for \vpec, indicating that the range of values
  for \Te\ obtained from different analyses does not significantly change the derived
  ICM constraints, particularly for \vpec.

  In addition, to determine the effect of assuming an isothermal
  ICM, rather than a multi-temperature ICM, a fit was performed
  based on the measurements presented in \citet{Johnson2012}.
  Specifically, they find two hot regions within the ICM
  with $T_{\mathrm{e}} \simeq 25 \pm 5$~keV.
  Based on the high--resolution SZ effect results presented in \citet{Mason2010},
  these hotter regions are likely to contribute $\lesssim 10$\% of the total
  SZ effect signal.
  If 90\% of the ICM is assumed to have $T_{\mathrm{e}} = 13.4 \pm 0.7$~keV,
  and 10\% is assumed to have $T_{\mathrm{e}} = 25 \pm 5$~keV, then the
  best fit values of \taue\ and \vpec\ are $6.55^{+0.87}_{-0.85} \times 10^{-3}$ and 
  $-1140^{+1060}_{-900}$~\kms, which differ by $0.78 \times 10^{-3}$ and 100~\kms\
  from the values obtained in the nominal isothermal fit.
  Particularly for \vpec, these differences are smaller than
  the measurement uncertainties, and indicate that the isothermal
  ICM assumption does not have a strong effect on the derived 
  \taue\ and \vpec\ constraints.

  As described in Section~\ref{sec:point_source}, a wide range of lower
  frequency observations were used to model the emission from the central
  AGN and to subtract it from the \bplusm\ data. To estimate the potential
  impact of the AGN emission on the final results, a fit was performed 
  without the AGN subtracted from the \bplusm\ images. The associated
  best-fit values of \taue\ and \vpec\ are $7.39^{+1.03}_{-1.02} \times 10^{-3}$ and 
  $-1910^{+840}_{-840}$~\kms, which differ from the nominal best-fit values by
  $0.06 \times 10^{-3}$ and 870~\kms. In the case of \vpec, this difference
  is significant, and indicates that the potential bias from the AGN
  emission is comparable to the measurement uncertainties.
  Analogously, a fit was performed without subtracting the \spire-derived
  model of the dusty star-forming galaxies, and the best-fit
  values of \taue\ and \vpec\ from this fit are $6.91^{+1.02}_{-1.04} \times 10^{-3}$
  and $-560^{+900}_{-890}$~\kms. These differ from the nominal best-fit values
  by $0.42 \times 10^{-3}$ and 480~\kms, and indicate that the
  potential \vpec\ bias from the dusty star-forming galaxy emission 
  is approximately 2 times lower than the potential bias from the AGN emission
  while the potential \taue\ bias is larger.
  
\section{Comparison to Previous Results}
  \label{sec:discussion}

  Our derived $v_{\mathrm{pec}} = -1040^{+870}_{-840}$~\kms\ is in 
  good agreement with the previously published 
  kinetic SZ effect value of $v_{\mathrm{pec}} = -1420^{+1270}_{-1170}$~\kms\
  from \citet{Kitayama2004}.
  Furthermore, both of these measurements are consistent with the
  expected distribution of the bulk component of \vpec\ for massive clusters,
  which has an rms of $\sim 200$--300~\kms (e.g., \citealt{Hernandez2010}).
  However, both results are noticeably offset from the value 
  of $v_{\mathrm{pec}} = +450 \pm 810$~\kms\ 
  derived in the kinetic SZ analysis of
  \citet{Zemcov2012}, although
  there are a number of systematic differences in measurement technique that could
  explain this difference.
  The \citet{Zemcov2012} measurement was made using the single-beam
  Z-Spec instrument and was therefore only sensitive to the value of \vpec\
  within a $\simeq 30''$ diameter region near the cluster center
  (compared to the more extended bulk cluster measurements
  in \citet{Kitayama2004} and in this work).
  In relatively small regions of the ICM, 
  similar to what was probed in the \citet{Zemcov2012} analysis,
  the internal ICM motions can be significant ($\simeq 500$~\kms, e.g., 
  \citealt{Nagai2003, Dolag2013}), and such motions could easily
  reconcile the results.
  In particular, merger activity near the center of \rxj, for which evidence
  is seen by \citet{Mason2010}, \citet{Johnson2012}, and \citet{Kreisch2014}, could
  potentially source such large internal motions in that region.

  \begin{deluxetable*}{cccc}
    \tablewidth{0pt}
     \tablecaption{SZ and X-ray Derived \mgas}
     \tablehead{\colhead{X-ray analysis} & \colhead{\rfive\ (Mpc)} & 
       \colhead{X-ray \mgas\ ($10^{13}$ M$_{\odot}$)} &
       \colhead{SZ \mgas\ ($10^{13}$ M$_{\odot}$)}}
     \startdata
       \citet{Czakon2015} & 0.71 & \phn$9.2 \pm 1.0$ & $11.9 \pm 1.7$ \\
       \citet{Donahue2014} & 0.74 & \phn$9.5 \pm 0.5$ & $11.7 \pm 1.8$ \\
       \citet{Mantz2014} & 0.80 & $13.6 \pm 1.5$ & $13.1 \pm 1.8$
     \enddata
     \tablecomments{Comparison of \mgas\ values obtained from three prior
       X-ray-only analyses to that obtained from this analysis incorporating
       SZ data. For each comparison, the SZ analysis is redone using
       the indicated X-ray-derived value of \rfive, though with a common
       \Te\ prior as described in Section~\ref{sec:X-ray}.
       As noted in the text, variations in \Te\ due to these small changes
       in aperture are negligible compared to the measurement uncertainty
       on \Te.
       For both the \citet{Czakon2015} and \citet{Donahue2014}
       comparisons, the SZ data find a value of \mgas\
       that is approximately $2\sigma$ larger than the X-ray value,
       while the agreement is good in the case of the \citet{Mantz2014}
       comparison.}
     \label{tab:mgas}
  \end{deluxetable*}

  Our derived $\tau_{\mathrm{e}} = 7.39^{+1.03}_{-1.02} \times 10^{-3}$,
  which gives the mean line-of-sight integrated electron optical depth within 
  the plane-of-sky aperture \rfive,
  can be used to obtain a constraint on the gas mass (\mgas) of \rxj.
  First, the parametric model given in Section~\ref{sec:model} is used to
  convert between \taue\ and $\tau_{\mathrm{e, sphere}} = 5.48 \pm 0.77 \times 10^{-3}$,
  where $\tau_{\mathrm{e, sphere}}$ represents the mean optical depth
  integrated within the line-of-sight defined by \rfive\ and the small asymmetry
  in the uncertainties on \taue\ is neglected.
  The gas mass can then be determined according to
  \begin{displaymath}
    M_{\textrm{gas}} = \left( \frac{\tau_{\mathrm{e, sphere}}}{\sigma_T} \right)
      \left(\pi R_{\mathrm{2500}}^2 \right) \left( \mu_{\mathrm{e}} m_{\textrm{p}} \right),
  \end{displaymath}
  where $\mu_{\mathrm{e}} = 1.14$ is the mean molecular weight per free electron
  and $m_{\textrm{p}}$ is the proton mass, yielding
  $M_{\textrm{gas}} = 11.9 \pm 1.7 \times 10^{13}$~M$_{\odot}$.
  We compare our SZ-derived \mgas\ to prior X-ray-derived \mgas\ values
  in Table~\ref{tab:mgas}, including the value of \mgas\ found by \citet{Czakon2015}
  as part of their X-ray analysis 
  to define the value of \rfive\ used throughout this manuscript. 
  In each case, the SZ estimation
  of \mgas\ was made using the \rfive\ value obtained by the indicated
  X-ray analysis. The prior on \Te\ was kept at its nominal value,
  as small changes in aperture have a negligible effect on \Te\ 
  relative to its uncertainty.

  This comparison indicates that, on average, the analyses incorporating SZ data
  yield systematically higher values of \mgas\ than the X-ray-only analyses.
  Such a difference could easily be explained by an elongation of the
  cluster along the line of sight, which would boost the SZ signal compared
  to the X-ray signal (e.g., \citealt{Morandi2012}).
  This result is in good agreement with \citet{Pointecouteau2001} and \citet{Kitayama2004},
  who also find a slight preference for line-of-sight elongation based on
  an SZ/X-ray comparison.
  In contrast, \citet{Chakrabarty2008}, \citet{Bonamente2012}, and \citet{Plagge2013}
  all find significantly lower SZ signals for \rxj\ than expected from
  the X-ray data, and those analyses imply a line-of-sight compression with axial ratios
  ranging from $\sim 3$--5.
  The reason for the disagreement between these two sets of results
  is not clear, although it may be related to
  differences in the SZ data. For example, \citet{Pointecouteau2001}, \citet{Kitayama2004},
  and this work use bolometric imaging data at mm wavelengths while
  \citet{Chakrabarty2008}, \citet{Bonamente2012}, and \citet{Plagge2013} use
  interferometric data at cm wavelengths.
  In addition, all six SZ datasets have sensitivities
  to different angular scales.
  Furthermore, the techniques used to fit the SZ and X-ray data, and to compare
  them, are different in all six of the analyses.
  Resolving these discrepancies is beyond the scope of this work
  and will likely require a dedicated, uniform
  re-analysis of all the relevant data.

\section{Summary}
  \label{sec:summary}

  By combining four-band MUSIC images with a 140~GHz Bolocam image we have 
  constrained the thermal and kinetic SZ effect signals towards \rxj.
  Based on these constraints, and including a prior on the 
  ICM temperature derived from X-ray spectroscopic
  measurements from \chandra,
  we find $v_{\mathrm{pec}} = -1040^{+870}_{-840}$~\kms.
  This value is in good agreement with the previously published result
  from \citet{Kitayama2004}, but is in mild tension with the velocity
  found by \citet{Zemcov2012}. However, some or all of that tension
  may be due to internal cluster motions within
  the relatively small cluster volume sampled by 
  \citet{Zemcov2012}.
  We find a mean electron optical depth within \rfive\ of
  $\tau_{\mathrm{e}} = 7.39^{+1.03}_{-1.02} \times 10^{-3}$, which
  implies a gas mass within \rfive\ of $M_{\textrm{gas}} = 11.9 \pm 1.7 \times 10^{13}$~M$_{\odot}$.
  This mass is slightly higher than previous values of \mgas\ derived
  from X-ray analyses, implying a slight elongation of \rxj\
  along the line of sight.

  For this analysis, we have included a rigorous treatment of the 
  photon and instrumental noise, noise due to 
  fluctuations in the atmospheric brightness, uncertainties in the
  planetary flux calibration model, and noise from three
  unwanted astrophysical sources (primary CMB fluctuations, dusty
  star-forming galaxies, and the cluster's central AGN). In the case
  of the dusty star-forming galaxies, we have used data from
  \spire\ to subtract the brightest objects.
  Most of the
  noise sources are correlated between the five observing bands, and
  all of these covariances have been accounted for in our overall results.
  For the four MUSIC bands, the total noise is largely due to
  the combination of photon, instrumental, and atmospheric noise,
  which implies that longer integration times would yield 
  significantly better constraints.
  For example, if we eliminate these noise sources from our analysis,
  then the uncertainties on \taue\ and \vpec\ are reduced by approximately
  a factor of two to $\pm 0.44 \times 10^{-3}$ and $\pm 460$~\kms, respectively.
  In addition, more advanced data-processing algorithms to remove
  the atmospheric fluctuations that are correlated between the 
  four MUSIC observing bands may produce noticeable improvements
  \citep{Adam2014, Benson2003}.

\section{Acknowledgments}

  We thank the anonymous referee for his or her extensive
  suggestions to improve this manuscript.
  We acknowledge the assistance of: the day crew and Hilo
  staff of the Caltech Submillimeter Observatory, who provided
  invaluable assistance during data-taking for this data set; 
  Kathy Deniston and Diana Bisel, who provided effective
  administrative support at Caltech and in Hilo;
  JS was partially supported by a Norris Foundation Fellowship;
  NGC was partially supported by a NASA Graduate Student Research
  Fellowship; SRS was supported by a NASA Earth and Space Science Fellowship;
  MUSIC was constructed and commissioned with funding provided by the Gordon and Betty Moore
  Foundation, JPL internal funds, and the NSF Advanced Technologies and Instrumentation (ATI)
  and Astronomy and Astrophysics Grants (AAG) Programs.
  This work is based in part on observations made with {\it Herschel},
  a European Space Agency Cornerstone Mission with a significant
  participation by NASA. Partial support for this work was provided
  by NASA through an award issued by JPL/Caltech.
  This research was performed while TM held a National Research Council
  Research Associateship Award at the Naval Research Laboratory.

  {\it Facilities:} \facility{Caltech Submillimeter Observatory}, \facility{{\it Chandra}}, \facility{{\it Herschel}}.  

\newpage
\bibliography{ms}
\bibliographystyle{apj}

\end{document}